\documentclass[a4paper,11pt]{article}

\usepackage{jheppub_mod} 
\usepackage[T1]{fontenc} 
\usepackage[utf8]{inputenc}

\usepackage{booktabs}
\usepackage{braket}
\usepackage[mathscr]{euscript} 
\usepackage{siunitx} 
\usepackage{mathtools} 

\usepackage{caption} 
\usepackage{subcaption}

\usepackage{pgfplots}
\pgfplotsset{compat=1.13} 

\usepackage[sort=none]{glossaries}
\glsdisablehyper

\DeclareUnicodeCharacter{2212}{-} 

\let\originalleft\left
\let\originalright\right
\renewcommand{\left}{\mathopen{}\mathclose\bgroup\originalleft}
\renewcommand{\right}{\aftergroup\egroup\originalright}

\newcommand{\Mrho}{M_{\rho}}
\newcommand{\Gammarho}{\Gamma_{\rho}}
\newcommand{\Mp}{M_{\pi}}
\newcommand{\Momega}{M_{\omega}}

\renewcommand{\hat}[1]{\widehat{#1}}

\newcommand{\mc}[1]{\mathcal{#1}}
\newcommand{\mb}[1]{\mathbb{#1}}
\newcommand{\mscr}[1]{\mathscr{#1}}

\renewcommand{\vec}[1]{\boldsymbol{#1}}
\newcommand{\abs}[1]{\left\vert#1\right\vert} 

\renewcommand{\Im}[0]{\text{Im}}
\renewcommand{\Re}[0]{\text{Re}}

\newcommand{\diff}{\text{d}} 
\DeclareMathOperator{\Disc}{disc} 
\DeclareMathOperator{\Res}{res} 
\newcommand{\GeV}{\,\text{GeV}}
\newcommand{\fm}{\,\text{fm}}
\DeclareMathOperator{\Sign}{sgn} 

\newcommand{\ROMAN}[1]{%
  \textup{\uppercase\expandafter{\romannumeral#1}}%
}

\newcommand{\Eqref}[1]{Eq.~\eqref{#1}}
\newcommand{\Figref}[1]{Fig.~\ref{#1}}
\newcommand{\Tabref}[1]{Table~\ref{#1}}
\newcommand{\Secref}[1]{Sec.~\ref{#1}}

\newacronym{chpt}{ChPT}{chiral perturbation theory}
\newacronym{cm}{CM}{center-of-mass}
\newacronym{hadspec}{HadSpec}{Hadron Spectrum Collaboration}
\newacronym{iam}{IAM}{inverse-amplitude method}
\newacronym{lec}{LEC}{low-energy constant}
\newacronym{lo}{LO}{leading-order}
\newacronym{nlo}{NLO}{next-to-leading-order}
\newacronym{nnlo}{NNLO}{next-to-next-to-leading-order}
\newacronym{qcd}{QCD}{quantum chromodynamics}
\newacronym{kt}{KT}{Khuri--Treiman}
\newacronym{irrep}{irrep}{irreducible representation}
\newacronym{vmd}{VMD}{vector-meson-dominance}

\title{The $\boldsymbol{\gamma\pi\to\pi\pi}$ anomaly from lattice QCD and dispersion relations}

\author[a]{Malwin Niehus,}
\author[b]{Martin Hoferichter,}
\author[a]{and Bastian Kubis}

\affiliation[a]{
Helmholtz-Institut f\"ur Strahlen- und Kernphysik (Theorie) and \\
Bethe Center for Theoretical Physics, Universit\"at Bonn, 53115 Bonn, Germany}

\affiliation[b]{
Albert Einstein Center for Fundamental Physics, Institute for Theoretical Physics, University of Bern, Sidlerstrasse 5, 3012 Bern, Switzerland}

\emailAdd{niehus@hiskp.uni-bonn.de}
\emailAdd{hoferichter@itp.unibe.ch}
\emailAdd{kubis@hiskp.uni-bonn.de}

\abstract{We propose a formalism to extract the $\gamma\pi\to\pi\pi$ chiral anomaly $F_{3\pi}$ from calculations in lattice QCD performed at larger-than-physical pion masses. To this end, we start from a dispersive representation of the $\gamma^{(*)}\pi\to\pi\pi$ amplitude, whose main quark-mass dependence arises from the $\pi\pi$ scattering phase shift and can be derived from chiral perturbation theory via the \acrlong{iam}. With parameters constrained by lattice calculations of the $P$-wave phase shift, we use this combination of dispersion relations and effective field theory to extrapolate two recent $\gamma^{(*)}\pi\to\pi\pi$ calculations in lattice QCD to the physical point. Our formalism allows us to extract the radiative coupling of the $\rho(770)$ meson and, for the first time, the chiral anomaly $F_{3\pi}=38(16)(11)\GeV^{-3}$. The result is consistent with the chiral prediction albeit within large uncertainties, which will improve in accordance with progress in future lattice-QCD computations.   
}


\begin{document} 
\maketitle
\glsresetall

\section{Introduction}\label{sec:introduction}
The calculation of hadronic scattering processes constitutes a notoriously challenging task in lattice \gls{qcd}, given the complications that ensue once multi-hadron dynamics are properly taken into account~\cite{Briceo2018}. In the past years, significant improvement has been achieved mainly for $\pi\pi\to\pi\pi$~\cite{Wilson2015,Bali:2015gji,Bulava:2016mks,Guo:2016zos,Briceno:2016mjc,Fu:2016itp,Alexandrou:2017mpi,Fu:2017apw,Andersen2019,Culver:2019qtx,Werner:2019hxc,Erben:2019nmx}, the simplest hadronic scattering process,  
 with computations even at physical pion mass available~\cite{Fischer:2020yvw,RBC:2021acc}. Meanwhile, the computation and analysis of more complicated processes remains challenging.
Already for the generalization when a pion is replaced by an external electromagnetic current, $\gamma^{(*)}\pi\to\pi\pi$,
up to now only two results are  available that correctly describe the resonant nature of the process~\cite{Briceno:2015dca,Briceno:2016kkp,Alexandrou:2018jbt}, both obtained at unphysically high pion masses exceeding \SI{300}{MeV}.
Hence, tools are needed to extrapolate the lattice-\gls{qcd} results to the physical point and confront them with or even improve upon experimental data.
In addition to defining an ideal test case to extend the lattice-\gls{qcd} calculation of elastic scattering to more complicated processes, phenomenological interest in $\gamma\pi\to\pi\pi$ itself motivates a detailed study of the extrapolation to the physical point.
At low energies, its form is dictated by the Wess--Zumino--Witten anomaly~\cite{Wess:1971yu,Witten:1983tw,Adler:1971nq,Terentev:1971cso,Aviv:1971hq}, leading to a theoretical prediction that has been tested experimentally at the \SI{10}{\percent} level~\cite{Antipov:1986tp,Giller:2005uy,Hoferichter:2012pm}, including the study of higher-order chiral corrections~\cite{Bijnens:1989ff,Holstein:1995qj,Hannah:2001ee,Ametller:2001yk,Bijnens:2012hf} and dispersive techniques~\cite{Hannah:2001ee,Truong:2001en,Hoferichter:2012pm,Hoferichter:2017ftn}---to be contrasted with the $\pi^0\to\gamma\gamma$ anomaly, whose chiral prediction, $F_{\pi\gamma\gamma}=1/(4\pi^2F_\pi)=0.2745(3)\GeV^{-1}$, has been confronted with experiment at sub-percent precision, $F_{\pi\gamma\gamma}=0.2754(21)\GeV^{-1}$~\cite{PrimEx-II:2020jwd}.
Furthermore, the process $\gamma\pi\to\pi\pi$ provides input to the data-driven Standard-Model prediction of the anomalous magnetic moment of the muon $a_\mu$, constraining both hadronic vacuum polarization and hadronic-light-by-light scattering. In view of the current $4.2\sigma$ discrepancy between the resulting prediction~\cite{Aoyama:2020ynm,Aoyama:2012wk,Aoyama:2019ryr,Czarnecki:2002nt,Gnendiger:2013pva,Davier:2017zfy,Keshavarzi:2018mgv,Colangelo:2018mtw,Hoferichter:2019gzf,Davier:2019can,Keshavarzi:2019abf,Hoid:2020xjs,Kurz:2014wya,Melnikov:2003xd,Colangelo:2014dfa,Colangelo:2014pva,Colangelo:2015ama,Masjuan:2017tvw,Colangelo:2017qdm,Colangelo:2017fiz,Hoferichter:2018dmo,Hoferichter:2018kwz,Gerardin:2019vio,Bijnens:2019ghy,Colangelo:2019lpu,Colangelo:2019uex,Blum:2019ugy,Colangelo:2014qya}\footnote{For more recent developments see, e.g., Refs.~\cite{Borsanyi:2020mff,Lehner:2020crt,Crivellin:2020zul,Achasov:2020iys,Keshavarzi:2020bfy,Malaescu:2020zuc,Colangelo:2020lcg} (hadronic vacuum polarization) and Refs.~\cite{Hoferichter:2020lap,Ludtke:2020moa,Bijnens:2020xnl,Bijnens:2021jqo,Zanke:2021wiq,Chao:2021tvp,Danilkin:2021icn,Colangelo:2021nkr} (hadronic light-by-light scattering).}
and experiment~\cite{Muong-2:2006rrc,Muong-2:2021ojo,Muong-2:2021vma,Muong-2:2021ovs,Muong-2:2021xzz}, further constraints on the hadronic amplitudes would of course be highly welcome.  
Phenomenologically, $\gamma\pi\to\pi\pi$ is dominated by the $\rho(770)$ resonance, thereby providing access to the radiative coupling of the $\rho$ to a photon and a pion~\cite{Hoferichter:2017ftn}, with the full kinematic dependence required to improve \gls{vmd} approaches. In particular, the amplitudes that enter the hadronic contributions to $a_\mu$ actually depend on the virtual process $\gamma^*\pi\to\pi\pi$, an extension that automatically arises in lattice \gls{qcd}.
Equivalently, the process is related to the decay $\gamma^*\to3\pi$ via crossing symmetry, and thus connected to the vector-meson decays $\omega(782)\to3\pi$ and $\phi(1020)\to3\pi$ when the virtuality of the photon coincides with the respective mass.
In this regard, the analysis of lattice-\gls{qcd} data in the scattering region provides a testing ground for frameworks that aim to analyze three-particle scattering directly, a subject that is currently under intense investigation~\cite{Hansen:2019nir,Rusetsky:2019gyk,Fischer:2020jzp,Hansen:2020otl,Mai:2021lwb}.
Accordingly, any model used to describe the $\gamma^{(*)}\pi\to\pi\pi$ lattice data needs to allow for a controlled extrapolation in the pion mass, work in the presence of a resonance, be accurate both in the low-energy region and in the complex plane where the $\rho$ pole is located, and, ideally, respect crossing symmetry and allow for a description of the decay region at the same time.
As such, \gls{chpt} by itself is insufficient, as unitarity is only restored perturbatively and resonances thus cannot be produced without unitarization. 
Instead, here we use a dispersive approach, based on the fundamental principles of unitarity and analyticity, which are implemented in the so-called \gls{kt} equations~\cite{Khuri:1960zz}. Their solution defines a set of reliable amplitudes for $\gamma\pi\to\pi\pi$ at the physical point~\cite{Hoferichter:2012pm,Hoferichter:2017ftn}, including the generalization to non-zero virtualities~\cite{Niecknig:2012sj,Danilkin:2014cra,Hoferichter:2014vra}. However, with the main input quantity the physical phase shifts for $\pi\pi$ scattering, this representation alone does not constrain the chiral extrapolation of lattice data, thus requiring the combination with \gls{chpt}. In particular, we use the \gls{iam}~\cite{Truong:1988zp,Dobado:1989qm,Truong:1991gv,Dobado:1992ha,Dobado:1996ps,Guerrero:1998ei,GomezNicola:2001as,Nieves:2001de} to describe $\pi\pi$ scattering at unphysical pion masses, as input for the solution of the \gls{kt} equations away from the physical point, based on the implementation from Ref.~\cite{Niehus:2020gmf}. In fact, the single-channel $SU(2)$ \gls{iam} can again be justified via a dispersion relation, with the only approximation regarding the chiral expansion of the left-hand cut~\cite{GomezNicola:2007qj}. The resulting expression can therefore not only be used to describe $\pi\pi$ scattering on the real axis, but also to study the pion-mass dependence of resonance trajectories~\cite{Hanhart:2008mx,Pelaez:2010fj} or form factors~\cite{Guo:2008nc,Colangelo:2021moe}.  
Following the strategy already laid out in Ref.~\cite{Niehus:2019nkl}, the combined \gls{kt} + \gls{iam}  representation for $\gamma^{(*)}\pi\to\pi\pi$
is then fit to lattice-\gls{qcd} data and afterwards extrapolated to the physical point, where the observables are extracted. At non-zero photon virtualities also the quark-mass dependence of the vector mesons $\omega$ and $\phi$ starts to play a role, see Ref.~\cite{Dax:2018rvs}, which can again be constrained using \gls{chpt} arguments~\cite{Jenkins:1995vb,Bijnens:1996nq,Bijnens:1997ni}.
This paper is organized as follows.
The basic form of the $\gamma^{(*)}\pi\to\pi\pi$ amplitude is introduced in \Secref{sec:basis}.
Subsequently, the dispersive framework used to analyze the lattice data and the fit procedure are discussed in \Secref{sec:dispersive} and \Secref{sec:fit}, respectively.
Finally, the fit results are presented in \Secref{sec:results} and conclusions drawn in \Secref{sec:conclusions}.

\section{The process \boldmath$\gamma\pi\to\pi\pi$}\label{sec:basis}
The scattering amplitude $\mc{M}$ of the process $\gamma^{(*)}(q)\pi^+(p)\to\pi^+(k)\pi^0(k^\prime)$ with four-momenta $q$, $p$, $k$, and $k^\prime$ can be expressed in terms of the electromagnetic current $J^\mu = e\big(\frac{2}{3} \overline{u}\gamma^\mu u - \frac{1}{3}\overline{d}\gamma^\mu d\big)$ as
\begin{equation}
  \mc{M}\left(s, t, q^2\right)
  =
  \epsilon^\mu\left(q^2\right)\Braket{\pi\pi,\vec{k},\vec{k}^\prime\left\vert J_\mu\left(0\right)\right\vert\pi,\vec{p}}.
\end{equation}
Here $e$ is the elementary charge, $\epsilon^\mu$ is the polarization vector of the photon, $s = (k + k^\prime)^2$ as well as $t = (p-k)^2$ denote the Mandelstam variables, and the pion states are normalized in the standard manner~\cite{Peskin1995}.
The pseudoscalar nature of the pions allows for decomposing the matrix element in terms of a complex-valued function $\mc{F}$ as
\begin{equation}\label{eq:matrix_element}
  \Braket{\pi\pi,\vec{k},\vec{k}^\prime\left\vert J_\mu\left(0\right)\right\vert\pi,\vec{p}}
  = i \epsilon_{\mu\nu\alpha\beta} p^\nu k^\alpha k^{\prime\beta}\mc{F}\left(s, t, q^2\right).
\end{equation}
At vanishing energy, the anomalous nature of the process fixes the amplitude in terms of the pion decay constant $F_\pi=\SI{92.28+-0.10}{\MeV}$~\cite{ParticleDataGroup:2020ssz} as $\mc{F}(0, 0, 0) = e F_{3\pi}$ with~\cite{Adler:1971nq} 
\begin{equation}\label{eq:anomaly}
  F_{3\pi}
  = \frac{1}{4\pi^2F_\pi^3}
  = \SI{32.23+-0.10}{\GeV^{-3}}.
\end{equation}
In particular, we follow the convention to absorb the class of chiral corrections that corresponds to the quark-mass renormalization of the decay constants of the three external pions into the physical $F_\pi$, and hence use \Eqref{eq:anomaly} as the reference point. Only chiral corrections that go beyond the quark-mass renormalization of $F_\pi$ will thus be applied in the matching~\eqref{matching}, see Refs.~\cite{Bijnens:1989ff,Hoferichter:2012pm}.

Throughout this work, isospin symmetry is assumed to hold.
In this case, only odd partial waves contribute, leading to the expansion~\cite{Jacob:1959at}
\begin{equation}
  \mc{F}\left(s, t, q^2\right)
  = \sum_{j=0}^{\infty}f_{2j+1}\left(s, q^2\right)P_{2j+1}^\prime\left(z\right).
\end{equation}
Here $f_J$ denotes the partial wave of total angular momentum $J=2j+1$, $z = \cos\theta$ with $\theta = \angle(\vec{k},\vec{k}^\prime)$ the scattering angle in the \gls{cm} system, and $P_{J}^\prime$ the derivative of the $J$-th Legendre polynomial.
Taking into account $\pi\pi$ intermediate states only, the $P$-wave fulfills the unitarity relation~\cite{Niecknig:2012sj}
\begin{equation}\label{eq:partial_wave_unitarity}
  \Im\left[f_1\left(s, q^2\right)\right]
  = f_1\left(s, q^2\right)\sigma_\pi\left(s\right)\big[T\left(s\right)\big]^*,
\end{equation}
for $s \geq 4\Mp^2$ with
\begin{equation}\label{eq:pion_pion_partial_wave_via_phase}
  T\left(s\right) = \frac{1}{\sigma_\pi\left(s\right)}\sin\left[\delta\left(s\right)\right]e^{i\delta\left(s\right)},
  \qquad \sigma_\pi\left(s\right) = \sqrt{1-\frac{4\Mp^2}{s}},
\end{equation}
the $\pi\pi\to\pi\pi$ $P$-wave amplitude and the $\pi\pi$ phase space, respectively, where $\delta = \arg[T]$ is the $P$-wave phase shift, $\Mp$ the pion mass, and the cut of the square-root is chosen along the positive real axis, which leads to $\sigma_\pi(s^*) = - \sigma_\pi(s)^*$.
The partial wave $T$ in turn obeys a slightly simpler unitarity relation, namely
\begin{equation}\label{eq:pion_pion_unitarity}
  \Im\left[T\left(s\right)\right]
  = \sigma_\pi\left(s\right)\abs{T\left(s\right)}^2,
\end{equation}
again for $s \geq 4\Mp^2$.
Equation~\eqref{eq:partial_wave_unitarity} implies (modulo $2\pi$)
\begin{equation}\label{eq:watsons_theorem}
  \arg\left[f_1\left(s,q^2\right)\right] =
  \begin{cases}
    \delta\left(s\right), & \Im\left[f_1\left(s, q^2\right)\right] \geq 0 \\
    \delta\left(s\right) - \pi, & \Im\left[f_1\left(s, q^2\right)\right] < 0
  \end{cases},
\end{equation}
which is a special case of Watson's theorem~\cite{Watson:1954uc}.
Building upon the unitarity relation~\eqref{eq:partial_wave_unitarity}, an expression for $f_1^{\ROMAN{2}}$, the $P$-wave on the second Riemann sheet, can be derived.
To that end, we make use of $f_1(s\pm i\epsilon,q^2) = f_1^{\ROMAN{2}}(s\mp i\epsilon, q^2)$ for $s \geq 4\Mp^2$ and $\epsilon\to 0$, the Schwarz reflection principle, i.e., $f_1(s^*) = f_1(s)^*$, as well as the uniqueness of analytic continuation to obtain
\begin{equation}\label{eq:partial_wave_second_sheet}
  f_1^{\ROMAN{2}}\left(s, q^2\right)
  = \frac{f_1\left(s, q^2\right)}{1 + 2i\sigma_\pi\left(s\right)T\left(s\right)}.
\end{equation}
Equation~\eqref{eq:partial_wave_second_sheet} exhibits a pole at $s_\rho = (M_\rho - i\Gamma_\rho/2)^2$ that is associated with the $\rho$ resonance of mass $M_\rho$ and width $\Gamma_\rho$ and accompanied by a twin pole at $s_\rho^*$.
The residue at $s_\rho$ factorizes into the coupling of the $\rho$ to the final state, $g_{\rho\pi\pi}$, as well as to the radiative coupling $g_{\rho\gamma\pi}$ to the initial state as~\cite{Hoferichter:2017ftn}
\begin{equation}\label{eq:partial_wave_residue}
  \Res\left[f_1^{\ROMAN{2}}, s_\rho\right]
  = -2 e g_{\rho\gamma\pi} g_{\rho\pi\pi}.
\end{equation}
Hence to determine the radiative coupling, $g_{\rho\pi\pi}$ needs to be known.
The latter can be fixed via
\begin{equation}
  \Res\left[T^{\ROMAN{2}}, s_\rho\right]
  = \frac{4\Mp^2-s_\rho}{48\pi}g_{\rho\pi\pi}^2,
\end{equation}
where
\begin{equation}\label{eq:pi_pi_second_sheet}
  T^{\ROMAN{2}}\left(s\right) = \frac{T\left(s\right)}{1 + 2i\sigma_\pi\left(s\right) T\left(s\right)}
\end{equation}
is the $\pi\pi$ $P$-wave on its second Riemann sheet, whose form is again dictated by unitarity, i.e., \Eqref{eq:pion_pion_unitarity}.
Note that to extract this couplings it is necessary to evaluate $T$ in the complex plane, hence a parameterization of $T$ is needed that goes beyond \Eqref{eq:pion_pion_partial_wave_via_phase}, which is valid only along the real axis above threshold.
In lattice-\gls{qcd} computations, the matrix element~\eqref{eq:matrix_element} is usually expressed in a different manner.
To that end, the $\pi\pi$ final state is expanded into components $\ket{\pi\pi,P,J,m}$ of total angular momentum $J$ and magnetic quantum number $m$.
The $P$-wave contribution to the matrix element is subsequently decomposed as~\cite{Briceno:2016kkp}
\begin{equation}\label{eq:matrix_element_lattice_convention}
  \Braket{\pi\pi,P,1,m\left\vert J_\mu\left(0\right)\right\vert\pi,\vec{p}}
= e \frac{2i}{\Mp} \epsilon_{\mu\nu\alpha\beta}p^\nu\epsilon^{*\alpha}\left(m,P\right)P^\beta \mc{A}\left(s, q^2\right),
\end{equation}
where the total momentum is given as $P = k + k^\prime$, $\epsilon^*(m,P)$ is the polarization vector of the two outgoing pions, i.e., the standard polarization vector of a spin 1 particle, and $\mc{A}$ is a complex-valued function.
In Ref.~\cite{Briceno:2016kkp} it is shown that $\mc{A}(s,q^2)\propto k_\text{CM}f_1(s,q^2)$.
Re-performing the computation, this time keeping track of all factors, results in
\begin{equation}\label{eq:amplitude_to_partial_wave}
  \abs{\mc{A}\left(s, q^2\right)}
  = \frac{\Mp k_\text{CM}}{2 e \sqrt{3}} \abs{f_1\left(s, q^2\right)},
\end{equation}
with $k_\text{CM}$ the absolute value of the momentum of a final-state pion in the \gls{cm} frame, i.e., $s = 4(\Mp^2 + k_\text{CM}^2)$.

For the remainder of this work, we will ignore all partial waves with $J \geq 3$.
The on-shell cross section is then given as~\cite{Hoferichter:2012pm,Briceno:2016kkp}
\begin{equation}\label{eq:cross_section}
  \sigma\left(s\right)
  = \frac{\left(s-4\Mp^2\right)^{3/2}\left(s-\Mp^2\right)}{768\pi\sqrt{s}}\abs{f_1\left(s,0\right)}^2.
\end{equation}

\section{Dispersive representation of \boldmath$\gamma\pi\to\pi\pi$}\label{sec:dispersive}
In the energy region of interest, where the $P$-wave dominates, $\gamma^{(*)}\pi\to\pi\pi$ can be accurately described by the \gls{kt} framework~\cite{Khuri:1960zz}.
Building upon dispersion relations and elastic unitarity~\eqref{eq:partial_wave_unitarity}, the \gls{kt} equations take into account not only individual $\pi\pi$ rescattering in the $s$-, $t$-, and $u$-channel, but also mixed rescattering, where pions rescatter, e.g., in the $t$-channel and subsequently in the $s$-channel.
The starting point is the reconstruction theorem~\cite{Hannah:2001ee},
\begin{equation}\label{eq:reconstruction_theorem}
  \mc{F}\left(s, t, q^2\right)
  = \mc{B}\left(s, q^2\right) + \mc{B}\left(t, q^2\right) + \mc{B}\left(u, q^2\right),
\end{equation}
which decomposes $\mc{F}$ into functions of a single Mandelstam variable only.
Here $u = 3\Mp^2 + q^2 - s - t$ is not a free variable.
The \gls{kt} equations then take the form~\cite{Niecknig:2012sj,Hoferichter:2012pm,Niehus:2019nkl}
\begin{equation}\label{eq:khuri_treiman}
  \begin{split}
    \mc{B}\left(s, q^2\right)
    &= \sum_{k=0}^{n-1} c_k\left(q^2\right)\mc{B}_k\left(s, q^2\right), \\
    \mc{B}_k\left(s, q^2\right)
    &= \Omega\left(s\right)\left[s^k + \frac{s^n}{\pi}\int\limits_{4\Mp^2}^\infty \frac{\sigma_\pi\left(x\right)}{x^n\left(x-s\right)}\frac{T\left(x\right)}{\Omega\left(x\right)}\hat{\mc{B}}_k\left(x, q^2\right)\diff x\right], \\
    \hat{\mc{B}}_k\left(s, q^2\right)
    &= \frac{3}{2}\int\limits_{-1}^1\left(1-z^2\right)\mc{B}_k\left(t\left(s, q^2, z\right), q^2\right)\diff z.
  \end{split}
\end{equation}
Here $n\in\mb{N}$ is the number of subtractions that are employed in the dispersive integrals, $c_k$ are the subtraction functions, the mappings $\mc{B}_k$ are known as basis functions, the Omn\`{e}s function is given as~\cite{Omnes:1958hv}
\begin{equation}
  \Omega\left(s\right)
  =\exp\left[\frac{s}{\pi}\int\limits_{4\Mp^2}^\infty\frac{\delta\left(x\right)}{x\left(x-s\right)}\diff x\right],
\end{equation}
and
\begin{equation}\label{eq:mandelstam_t}
\begin{split}
  t\left(s, q^2, z\right)
  &= \tau\left(s, q^2\right) + z \kappa\left(s, q^2\right), \\
\tau\left(s, q^2\right) &= \frac{3\Mp^2 + q^2 - s}{2}, \\
\kappa\left(s, q^2\right) &= \frac{1}{2}\sigma_\pi\left(s\right)\sqrt{\lambda\left(s, q^2, \Mp^2\right)},
\end{split}
\end{equation}
is the Mandelstam variable $t$ expressed in terms of the other kinematic variables via the K\"{a}ll\'{e}n function $\lambda(a, b, c) = (a - b - c)^2 - 4bc$.
The basis functions subsume the $\pi\pi$ rescattering and are fixed as soon as the $\pi\pi$ phase $\delta$ is known.
While the Omn\`{e}s function describes $\pi\pi$ scattering in one channel, the integral in \Eqref{eq:khuri_treiman} incorporates mixed rescattering.
That is, the replacement $\mc{B}_k(s,q^2) \mapsto s^k\Omega(s)$ amounts to taking into account only $\pi\pi$ rescattering in the individual channels.
In the form of \Eqref{eq:khuri_treiman} the \gls{kt} equations are valid only if $q^2 < (3\Mp)^2$, i.e., as long as the photon cannot decay.
By deformation of either one of the integration contours the equations can be analytically continued towards $q^2 > (3\Mp)^2$~\cite{Bronzan:1963mby,Gasser:2018qtg}, however, this is not needed for the lattice data of interest.
This analytic continuation reveals that the basis functions indeed possess a three-particle cut in $q^2$ that is associated with pairwise $\pi\pi$ rescattering, but they do not contain any $q^2$-dependence arising from genuine three-pion interactions~\cite{Aitchison:1966lpz}.
Such interactions are to be described by the subtraction functions $c_k$, which are not fixed by the \gls{kt} approach.
Thus, to arrive at a complete representation of $\gamma^{(*)}\pi\to\pi\pi$, we need both a representation of $\delta$ as well as a parameterization of the subtraction functions.
For the former, we employ the \gls{iam}.
In this approach, the $\pi\pi$ $P$-wave is expanded in $SU(2)$ \gls{chpt}, $T = T_2 + T_4 + \dots$, where $T_2$ denotes the \gls{lo} \gls{chpt} expression and $T_4$ the \gls{nlo} one.
This expansion satisfies \Eqref{eq:pion_pion_unitarity} only perturbatively, but we can unitarize it to obtain
\begin{equation}\label{eq:nlo_iam}
  T
  = \frac{T_2^2}{T_2 - T_4},
\end{equation}
which is precisely the \gls{nlo} \gls{iam}~\cite{Dobado:1989qm,Truong:1991gv,Dobado:1992ha}.
Equation~\eqref{eq:nlo_iam} satisfies \Eqref{eq:pion_pion_unitarity} exactly, exhibits the correct analytic structure, and is valid in the entire complex plane.
Explicit expressions for the \gls{chpt} amplitudes in closed analytical form are given, e.g., in Ref.~\cite{Niehus:2020gmf}, building upon the computations presented in Refs.~\cite{Weinberg:1966kf, Gasser1984}.
In addition to the pion mass $\Mp$, the amplitudes depend on $F$, the pion decay constant in the chiral limit, as well a $l^r = l_2^r - 2l_1^r$, a single linear combination of the ordinarily renormalized \glspl{lec} $l_1^r$, $l_2^r$.
It is beneficial to express the amplitudes in terms of $F$ instead of $F_\pi$, for the former is pion-mass independent, see also the discussion in Ref.~\cite{Niehus:2020gmf}.
In this work we use the $N_\text{f} = 2 + 1$ FLAG average of $F_\pi/F$~\cite{Aoki:2019cca, Bazavov:2010hj, Beane:2011zm, Borsanyi:2012zv, Durr:2013goa, Boyle:2015exm}, which yields $F=\SI{86.89(58)}{\MeV}$ when combined with the PDG value of $F_\pi$.
Unfortunately, the phase of the \gls{nlo} \gls{iam} does not approach $\pi$, instead, $\lim_{s\to\infty}T(s) = - [96\pi l^r + i + 2/(3\pi)]^{-1}$, which yields $\lim_{s\to\infty}\delta(s) < \pi$ for all reasonable values of $l^r$.
However, for the numerical computation of $\mc{B}_k$ it is beneficial if $\delta$ approaches $\pi$ at a finite value $\Lambda$ of Mandelstam $s$, since this provides a natural cutoff of the integral in \Eqref{eq:khuri_treiman} via $T(\Lambda) = 0$.
For this reason, and because the \gls{nlo} \gls{iam} is physically reasonable in the elastic region only, we guide $\delta$ smoothly to $\pi$ at energies far above the resonance region.
To be precise, we use the IAM below $s = 270\Mp^2$, $\delta(s) = \pi$ for $s \geq \Lambda = 310\Mp^2$, and a fourth-order polynomial in between such that the transition is smooth.
The suppression of the high-energy region due to the subtractions ensures that the systematic error introduced in this way is negligible compared to the error of the $\gamma\pi$ data, this suppression is particularly strong since we use not only one, but $n=2$ subtractions, see Sec.~\ref{sec:results}.
The basis functions can be computed numerically via standard methods~\cite{Gasser:2018qtg}.
The subtraction functions need to be holomorphic in the complex $q^2$-plane except for a cut along $[9\Mp^2, \infty)$ that is associated with $\gamma^*\to3\pi$.
Hence we can write down an $m$-times subtracted dispersion relation of the form
\begin{equation}\label{eq:subtractions_dispersive_integral}
  c_k\left(q^2\right)
  = \sum_{j=0}^{m-1} b_{kj} \left(q^2\right)^j + \frac{\left(q^2\right)^m}{2\pi i}\int\limits_{9\Mp^2}^\infty\frac{\Disc\left[c_k\left(x\right)\right]}{x^m\left(x-q^2\right)}\diff x,
\end{equation}
where $\Disc[c_k(x)] = \lim_{\epsilon\to 0}[c_k(x+i\epsilon) - c_k(x-i\epsilon)]$ denotes the discontinuity along the branch cut.
As long as $q^2 < 9\Mp^2$, the Schwarz reflection principle dictates $b_{kj}\in\mb{R}$ and $\Disc[c_k] = 2i\Im[c_k]$.
In the energy region that contributes most to the dispersive integral in \Eqref{eq:subtractions_dispersive_integral} the three-pion physics is dominated by the $\omega (782)$ and $\phi (1020)$ resonances~\cite{ParticleDataGroup:2020ssz}, both of which are narrow and (at the physical point) far away from the three-pion threshold.
Thus $\Disc[c_k]$ inside the integral can be reasonably well described by a sum of two Breit--Wigner functions, yielding a dispersively improved variant of a Breit--Wigner parameterization that ensures the correct analytic properties~\cite{Moussallam:2013una,Hoferichter:2014vra}.
In practice, the lattice data we are going to analyze are obtained at $\Mp > \SI{300}{\MeV}$, at which mass the $\omega$ becomes a bound state~\cite{Dax:2018rvs}.
Accordingly, instead of being incorporated into the dispersive integral, it appears as a pole at $q^2 = \Momega^2$.
This can be taken into account by writing down a dispersion relation in the form of \Eqref{eq:subtractions_dispersive_integral} for $c_k(q^2) / \mscr{P}(q^2)$ with $\mscr{P}(q^2) = (1 - q^2/\Momega^2)^{-1}$ and multiplying the result by the pole factor $\mscr{P}$.
Since the lattice data are obtained at virtualities significantly below the $3\pi$ threshold, \Eqref{eq:subtractions_dispersive_integral} can be expanded as a Taylor series, keeping the first $m$ terms yields
\begin{equation}\label{eq:polynomial}
  c_k\left(q^2\right) = \sum_{j=0}^{m-1} b_{kj} \left(q^2\right)^j.
\end{equation}
However, the convergence of the Taylor series is poor as soon as $\abs{q^2}$ gets close to the $3\pi$ threshold, this drawback goes hand in hand with a wrong asymptotic behavior for large $\abs{q^2}$, i.e., the expression diverges.
To improve on \Eqref{eq:polynomial}, a conformal polynomial can be used instead~\cite{Yndurain:2002ud}.
That is, \Eqref{eq:subtractions_dispersive_integral} is approximated by
\begin{equation}\label{eq:conformal_polynomial}
  c_k\left(q^2\right)
  = \sum_{j=0}^{m-1} b_{kj} w\left(q^2\right)^j,
\end{equation}
where the conformal variable $w$ reads
\begin{equation}
  w\left(q^2\right)
  = \frac{\sqrt{9M_\pi^2 - q^2} - 3M_\pi}{\sqrt{9M_\pi^2 - q^2} + 3M_\pi}.
\end{equation}
In this way, the cut along $[9\Mp^2, \infty)$ is retained, moreover, the asymptotic behavior is improved, as $w$ is bounded.
At the $3\pi$ threshold, $\Im[c_k]$ should scale like $(q^2 - 9\Mp^2)^4$ to be in accordance with the three-particle phase space~\cite{Leutwyler:2002hm}.
This is impossible to obtain with \Eqref{eq:conformal_polynomial}, because for $q^2$ below threshold the Schwarz reflection principle needs to be fulfilled, hence $b_{kj}\in\mb{R}$.
Expanding $w$ in powers of $x \coloneqq \sqrt{9\Mp^2 - q^2}$ makes it clear that only odd powers of $x$ contribute to $\Im[c_k]$.
This problem is fundamental to the method, for the Riemann mapping theorem implies that each biholomorphic map from the cut complex plane to the interior of the unit disc is of conformal form.
It is however easily possible to remove the leading square-root-like scaling via fixing~\cite{Colangelo:2018mtw}
\begin{equation}\label{eq:modified_threshold}
  b_{k1} = - \sum_{j=2}^{m-1} j b_{kj}.
\end{equation}
Altogether we have six different parameterizations of the subtraction functions: a polynomial, a conformal polynomial, and a conformal polynomial with modified threshold behavior, each either with or without the pole factor $\mscr{P}$ in front.
To take into account the pion-mass dependence of $\Momega$ appearing in $\mscr{P}$, we use the result of the analysis in Ref.~\cite{Dax:2018rvs}, namely
\begin{equation}
  \Momega\left(\Mp^2\right)
  = \SI{0.7686+-0.002}{\GeV} + \SI{0.719+-0.009}{\GeV^{-1}}\Mp^2.
\end{equation}
Lastly, the number of terms $m$ needs to be fixed.
We use $N - k$ terms with a single global value of $N$ for the $k$-th subtraction function $c_k$, since it multiplies $s^k$ in \Eqref{eq:khuri_treiman}, such that with our choice in the simple polynomial representation~\eqref{eq:polynomial} the highest combined power of Mandelstam $s$ and $q^2$ has mass dimension $2N$.
An overview of the different strategies is given in \Tabref{tab:strategies}.

\begin{table}
  \begin{center}
  \begin{tabular}{lccc}\toprule
  & \Eqref{eq:polynomial} & \Eqref{eq:conformal_polynomial} & \Eqref{eq:conformal_polynomial} and \Eqref{eq:modified_threshold} \\
  \midrule
    without $\mscr{P}$ & \ROMAN{1} & \ROMAN{2} & \ROMAN{3} \\
    with $\mscr{P}$ & $\ROMAN{1}\mscr{P}$ & $\ROMAN{2}\mscr{P}$ & $\ROMAN{3}\mscr{P}$ \\\bottomrule
  \end{tabular}
  \caption{The naming scheme of the different parameterizations of the subtraction functions.
    For example, strategy $\ROMAN{2}\mscr{P}$ amounts to $c_k(q^2) = \mscr{P}(q^2)\sum_{j=0}^{N-k}b_{kj}w(q^2)^j$.
  }\label{tab:strategies}
  \end{center}
\end{table}

Given the basis functions, the $P$-wave can be expressed as
\begin{equation}\label{eq:partial_wave}
  f_1\left(s, q^2\right)
  = \sum_{k=0}^{n-1} c_k\left(q^2\right)\left[\mc{B}_k\left(s, q^2\right) + \hat{\mc{B}}_k\left(s, q^2\right)\right].
\end{equation}
Hence it requires the computation of  $\hat{\mc{B}}_k$.
Doing so directly from its definition in \Eqref{eq:khuri_treiman} is in this context inconvenient for two reasons.
First, the lattice data contain several data points at $q^2 > \Mp^2$.
At these virtualities, Mandelstam $t$ develops a non-vanishing imaginary part due to the square root of the K\"{a}ll\'{e}n function in \Eqref{eq:mandelstam_t}.
Thus $\mc{B}_k$ needs to be evaluated at different lines in the complex plane, one for each $q^2 > \Mp^2$.
Second, we aim for an evaluation of $f_1$ at the resonance pole to extract the radiative coupling by means of \Eqref{eq:partial_wave_residue}.
Again, this requires the computation of $\mc{B}_k$ at complex values of Mandelstam $t$, where care needs to be taken to avoid collisions with the branch cut of $\mc{B}_k$.
To circumvent these issues, we resort to the kernel method~\cite{Hoferichter:2017ftn}, which allows for evaluation of $\hat{\mc{B}}_k$ at arbitrary values of Mandelstam $s$ by computing $\mc{B}_k$ along the real axis only.
The details are given in App.~\ref{sec:kernel_method}.
Altogether, the free parameters of our dispersive representation are $l^r$ as well as the variables $b_{kj}$ appearing in \Eqref{eq:polynomial} or \Eqref{eq:conformal_polynomial}.
While the former is pion-mass independent, the latter depend on $\Mp$.
Since there are only lattice data sets at two different pion masses available, only very simple parameterizations of the pion-mass dependence of each $b_{kj}$ can presently be constrained.
For that reason, we opt for the simplest ansatz
\begin{equation}\label{eq:subtraction_mass_dependence}
  b_{kj}\left(M_\pi^2\right)
  = \alpha_{kj} + \beta_{kj}M_\pi^2, \qquad \alpha_{kj}, \beta_{kj}\in\mb{R}.
\end{equation}
Here the fact that the variables are linear in $\Mp^2$ instead of $\Mp$ is motivated by \gls{chpt}, for otherwise the $b_{kj}$ would possess branch points in the quark mass.
As soon as more data sets at different pion masses become available, it will be possible to test more refined prescriptions.
With the subtraction functions extrapolated to the physical point, defined by the PDG value of the mass of the charged pion~\cite{ParticleDataGroup:2020ssz}, the anomaly~\eqref{eq:anomaly} can be determined via matching the dispersive representation to \gls{chpt}.
For $n=2$ subtractions the matching yields~\cite{Hoferichter:2012pm}
\begin{equation}
\label{matching}
  eF_{3\pi}\left(1 + G\right)
  = 3\left\{c_0\left(0\right)\left[1 + \frac{\diff\Omega}{\diff s}\left(0\right)\Mp^2\right] + c_1\left(0\right)\Mp^2\right\},
\end{equation}
with
\begin{equation}
  G
  = \frac{3}{2}\frac{\Mp^2}{\Mrho^2} - \frac{1}{32\pi^2}\frac{\Mp^2}{F^2}\left[1 + \log\frac{\Mp^2}{\Mrho^2}\right]
\end{equation}
determined via one-loop \gls{chpt} and a new \gls{lec} has been fixed via resonance saturation~\cite{Bijnens:1989ff}.

\section{Fit to lattice data}\label{sec:fit}
The lattice-\gls{qcd} computations of $\gamma^{(*)}\pi\to\pi\pi$ are based on the formalism presented in Ref.~\cite{Briceno:2014uqa}, which describes a two-step approach.
First, $N_{\pi\pi}$ different $\pi\pi$ \gls{cm} energy levels $E^\text{lat}_k$, $k=1,\dots,N_{\pi\pi}$ are computed in the finite volume, which are related to the phase shift $\delta$ via L\"uscher's quantization condition~\cite{Luscher:1990ux,Briceo2018},
\begin{equation}\label{eq:luescher_quantization_condition}
  0 = \left[\delta\left(s\right) - \mscr{Z}\left(s\right)\right]_{\sqrt{s} = E^\text{lat}_k}.
\end{equation}
Here $\mscr{Z}$ is a known expression that depends on the kinematics and the characteristics of the lattice.
Second, a finite-volume version of the matrix element~\eqref{eq:matrix_element_lattice_convention} is computed, from which one can extract $\mc{A}_\text{FV}$, the finite-volume analog of $\mc{A}$ in \Eqref{eq:matrix_element_lattice_convention}.
The latter is related to its finite-volume counterpart by~\cite{Briceno:2014uqa, Briceno:2016kkp}
\begin{equation}\label{eq:finite_volume_to_infinite_volume_amplitude}
  \abs{\mc{A}\left(s, q^2\right)}^2
  = \mscr{L}\left(s\right)\abs{\mc{A}_\text{FV}\left(s, q^2\right)}^2,
\end{equation}
where the Lellouch--L\"uscher factors are given as\footnote{These factors yield the $\gamma^{(*)}\pi^+\to\pi^+\pi^0$ amplitude~\cite{Briceno:2014uqa}. They differ from the ones in Refs.~\cite{Briceno:2016kkp,Alexandrou:2018jbt} by a factor $2$, because the formulae given in both references apply to the isospin-projected $\gamma^{(*)}\pi\to\pi\pi(I=1)$ amplitude~\cite{Briceno:2021}.  
We thank Ra\'ul Brice\~no and Marcus Petschlies for extensive discussions on this point.}
\begin{equation}
  \mscr{L}\left(s\right)
  = \frac{4\pi}{k_\text{CM}}\frac{\partial}{\partial \sqrt{s}}\left[\delta\left(s\right) - \mscr{Z}\left(s\right)\right].
\end{equation}
These factors are uniquely defined only on the energies that are solutions of \Eqref{eq:luescher_quantization_condition}, that is, they are not defined as functions of arbitrary values of $s$~\cite{Briceno:2021dzc}.
Notably, the computation of $\mscr{L}$ requires the computation of the derivative of the phase shift.
Since the lattice data points are too sparse, it is not feasible to compute the derivative by an interpolation of the data.
Instead, a continuous parameterization is needed, we use the \gls{nlo} \gls{iam} as given in \Eqref{eq:nlo_iam}.
Accordingly, the fit to the data works as follows.
First, the \gls{nlo} \gls{iam} is fit to the $\pi\pi$ energy levels by minimizing
\begin{equation}\label{eq:chi_2_pi_pi}
  \chi^2_{\pi\pi}\left(l^r\right)
  = \sum_{j, k=1}^{N_{\pi\pi}} \left[E^\text{lat}_j - E^\text{IAM}_j\left(l^r\right)\right]
  \left[C^{-1}_{\pi\pi}\right]_{jk}\left[E^\text{lat}_k - E^\text{IAM}_k\left(l^r\right)\right]
\end{equation}
with respect to the fit parameter $l^r$.
Here $C_{\pi\pi}$ is the covariance matrix of the lattice $\pi\pi$ energies $E_k^\text{lat}$, and $E_k^\text{IAM}(l^r)$ is obtained by substituting $\delta$ in \Eqref{eq:luescher_quantization_condition} with the \gls{iam} phase shift and solving the equation for $\sqrt{s} = E_k^\text{IAM}$, with the kinematics of the $k$th lattice energy level, as explained in detail in Ref.~\cite{Niehus:2020gmf}.
Second, the derivative of the \gls{iam} phase for the resulting value of $l^r$ is used to compute the Lellouch--L\"uscher factors.
To be consistent, the factors are evaluated at the energies $E_k^\text{IAM}$, for these are the solutions of L\"uscher's quantization condition with the \gls{iam} phase shift.
Inserting the factors obtained this way into \Eqref{eq:finite_volume_to_infinite_volume_amplitude} and combining it with \Eqref{eq:amplitude_to_partial_wave} allows us to compute
\begin{equation}\label{eq:partial_wave_from_lattice}
  \abs{f_1^\text{lat}\left(s_k^\text{lat}, q^{2\,\text{lat}}_a\right)}
  = \frac{2e\sqrt{3}}{\Mp {k_\text{CM}}_k}\sqrt{\mscr{L}\left(s_k^\text{IAM}\right)}\abs{\mc{A}_\text{FV}\left(s_k^\text{lat}, q^{2\,\text{lat}}_a\right)}
  ,
\end{equation}
the absolute value of the partial wave, with $s_k^\text{lat} = (E_k^\text{lat})^2$, $s_k^\text{IAM} = (E_k^\text{IAM})^2$, and $q_a^{2\,\text{lat}}$ the virtuality of the corresponding lattice data point.
The difference $\abs{E_k^\text{IAM} - E_k^\text{lat}}$ is small, but the \gls{iam} energies do not agree perfectly with the lattice ones.
While at a first sight the appearance of two different energies in \Eqref{eq:partial_wave_from_lattice} might seem problematic, at the current level of precision of the data it is irrelevant.
Third, the resulting values of $\abs{f_1^\text{lat}}$ are used to fit the $P$-wave as computed via \gls{kt} equations.
The lattice-\gls{qcd} computation yields $N_{\gamma\pi}$ values of $\mc{A}_\text{FV}$, corresponding to $N_{\gamma\pi}$ different virtualities $q^{2\,\text{lat}}_a$, $a=1,\dots,N_{\gamma\pi}$, at $N_{\pi\pi}$ different energies, all of which have errors.
For the data sets at hand, $N_{\pi\pi} < N_{\gamma\pi}$, i.e., several data points are obtained at the same energy.
To take into account the errors of the energies and virtualities, we follow the standard approach and introduce an auxiliary fit parameter for each kinematic variable, see, e.g., Ref.~\cite{Alexandrou:2017mpi}, leading to
\begin{equation}\label{eq:chi_2_gamma_pi}
  \chi^2_{\gamma\pi}
  = \left(v^\text{lat} - v^\text{KT}\right) C_{\gamma\pi}^{-1}\left(v^\text{lat} - v^\text{KT}\right)^T
\end{equation}
with
\begin{equation}
  \begin{split}
  v^\text{lat}
  &= \begin{pmatrix}
    \abs{f_1^\text{lat}\left(s_1^\text{lat}, q^{2\,\text{lat}}_1\right)}, &
    \dots, &
    \abs{f_1^\text{lat}\left(s_{N_{\pi\pi}}^\text{lat}, q^{2\,\text{lat}}_{N_{\gamma\pi}}\right)}, &
    q^{2\,\text{lat}}_1, &
    \dots, &
    q^{2\,\text{lat}}_{N_{\gamma\pi}}, &
    E_1^\text{lat}, &
    \dots, &
    E_{N_{\pi\pi}}^\text{lat}
  \end{pmatrix}, \\
  v^\text{KT}
  &= \begin{pmatrix}
    \abs{f_1\left(s_1, q^{2}_1\right)}, &
    \dots, &
    \abs{f_1\left(s_{N_{\pi\pi}}, q^{2}_{N_{\gamma\pi}}\right)}, &
    q^{2}_1, &
    \dots, &
    q^{2}_{N_{\gamma\pi}}, &
    E_1, &
    \dots, &
    E_{N_{\pi\pi}}
  \end{pmatrix},
  \end{split}
\end{equation}
the auxiliary fit parameters $q_1^2, \dots, q_{N_{\gamma\pi}}^2$, $E_1, \dots, E_{N_{\pi\pi}}$, as well as $s_k = E_k^2$, and covariance matrix $C_{\gamma\pi}$.
The error of the \gls{iam} phase leads to an error of the Lellouch--L\"uscher factors.
The corresponding covariance matrix is added to the appropriate entries of $C_{\gamma\pi}$.
Equation~\eqref{eq:chi_2_gamma_pi} is minimized with respect to the auxiliary fit parameters and the variables $b_{kj}$ appearing in the parameterization of the subtraction functions.
Since only the absolute value of the partial wave is fit and $f_1$ is linear in the fit parameters $b_{kj}$, the latter are fixed by the fit only up to a global phase $\pm 1$.
To fix this, we impose the upper case of \Eqref{eq:watsons_theorem}, i.e., $\arg[f_1(s, q^2)] = \delta(s)$.
There are different sources of error that need to be taken into account.
The $\pi\pi$ energy levels $E_k^\text{lat}$ carry an error due to the statistical nature of the lattice computation, this error is taken into account by jackknife resampling.
Furthermore, on the lattice everything is computed in units of the lattice spacing $a$.
The translation into physical units requires the determination of $a$, the so-called scale setting.
The resulting value of $a$ carries a statistical uncertainty, moreover, a systematic error arises for the scale setting is not unique away from the physical point.
To keep the impact of the scale setting and the associated error of $a$ as small as possible, we phrase both \Eqref{eq:chi_2_pi_pi} and \Eqref{eq:chi_2_gamma_pi} in lattice units.
However, in one place at the $\chi^2$-level the lattice spacing enters, namely via the decay constant $F$, whose literature value is required for the evaluation of the \gls{iam} amplitudes and needs to be translated into lattice units.
To assess the impact of the statistical error of the lattice spacing on the $\pi\pi$ fit, we perform a parametric bootstrap, see Ref.~\cite{Niehus:2020gmf} for further details.
We do not attempt to estimate the uncertainty associated with the systematic error of the lattice spacing.
To determine the error of the fit parameters of the $\gamma\pi$ fit, we simply use the Hessian.
In principle, the error of the \gls{iam} phase impacts the $\gamma\pi$ fit not only via the covariance matrix $C_{\gamma\pi}$, but also via the \gls{kt} equations.
In practice, the error of the phase is negligible compared to the error of $\abs{\mc{A}_\text{FV}}$.

\begin{table}
  \begin{center}
  \begin{tabular}{lllll}
    \toprule
    & fit & Ref.~\cite{Bijnens:2014lea} & FLAG~\cite{Aoki:2019cca} & Ref.~\cite{GarciaMartin:2011jx} \\
    \midrule
    $\chi^2/\text{dof}$ & $31.7 / (27 - 1) = 1.22$ \\
    $p$-value & $0.20$ \\
    $l^r \times 10^3$ & $12.79(11)(10)(12)$ & $9.9(1.3)$ & $19(17)$\\
    $\Mrho/\si{\MeV}$ & $747.2(2.7)(2.8)(1.0)$ & & & $763.7^{+1.7}_{-1.5}$\\
    $\Gammarho/\si{\MeV}$ & $145.0(1.9)(1.9)(1.0)$ & & & $146.4^{+2.0}_{-2.2}$ \\
    $\Re(g_{\rho\pi\pi})$ & $5.960(22)(21)(25)$ & & & $5.98^{+0.04}_{-0.07}$\\
    $-\Im(g_{\rho\pi\pi})$ & $0.7175(82)(79)(94)$ & & & $0.56^{+0.07}_{-0.10}$\\
    \bottomrule
  \end{tabular}
  \caption{The outcomes of the \gls{iam} fit to the $\pi\pi$ data.
  The first error arises due to the statistical error of the $\pi\pi$ data, the second due to the error of the lattice spacings, and the third due to the error of the literature value of $F$.
  The third and fourth column contain reference values for the \gls{lec} from \gls{chpt} and lattice \gls{qcd}~\cite{Aoki:2019cca, Bazavov:2010hj, Beane:2011zm, Borsanyi:2012zv, Durr:2013goa, Boyle:2015exm}, respectively, while the fifth column lists the $\rho$ properties as determined via Roy-like equations.
}\label{table:pi_pi_fit}
  \end{center}
\end{table}

\begin{figure}
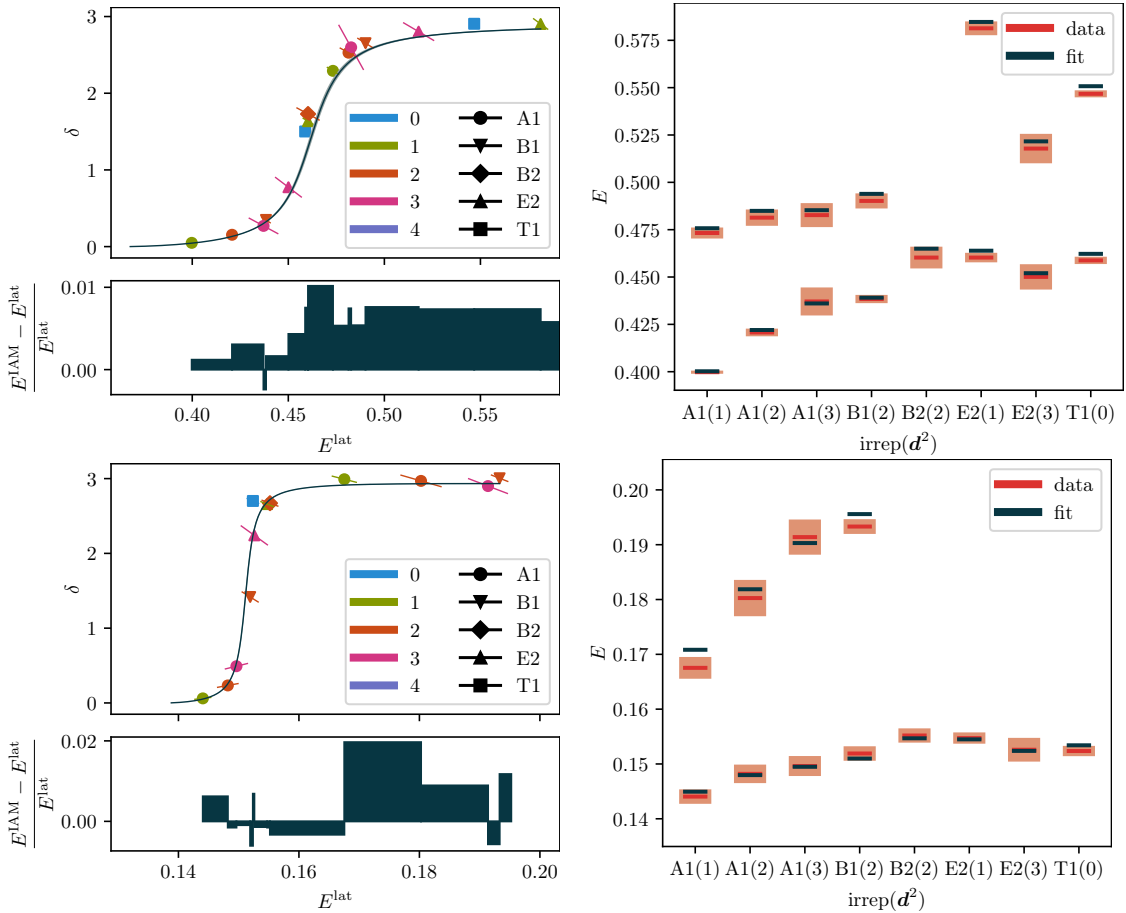

  \begin{subfigure}{0.495\textwidth}
    \scalebox{0.75}{\input{img/phase_data_320.pgf}}
  \end{subfigure}
  \begin{subfigure}{0.495\textwidth}
    \scalebox{0.75}{\input{img/energy_data_320.pgf}}
  \end{subfigure}
  \begin{subfigure}{0.495\textwidth}
    \scalebox{0.75}{\input{img/phase_data_390.pgf}}
  \end{subfigure}
  \begin{subfigure}{0.495\textwidth}
    \scalebox{0.75}{\input{img/energy_data_390.pgf}}
  \end{subfigure}
  \begin{center}
  \caption{The fit \gls{iam} in comparison to the $2\pi$ lattice data.
  The top (bottom) plots correspond to the lattice data of Ref.~\cite{Alexandrou:2017mpi} (Ref.~\cite{Dudek2013}). On the left-hand side, the phase is depicted. Here the markers encode the \glspl{irrep} of the residual rotational symmetry on the lattice, while the colors encode the square of the boost momentum $\vec{d}= \frac{L}{2\pi}\vec{P}$ with $L^3$ the spatial volume of the lattice. On the right-hand side, the comparison is shown on the energy level, with the statistical uncertainty of the lattice data indicated by the light red boxes. All energies are given in lattice units.}\label{fig:iam_fit}
  \end{center}
\end{figure}

\section{Results}\label{sec:results}
\subsection[Fits to $\pi\pi$ data]{Fits to $\boldsymbol{\pi\pi}$ data}
As already stressed, currently there are only two $\gamma\pi$ data sets available, one at $\Mp\approx\SI{317}{\MeV}$~\cite{Alexandrou:2018jbt}, the other one at $\Mp\approx\SI{391}{\MeV}$~\cite{Briceno:2015dca,Briceno:2016kkp}.
Hence, to fix the pion-mass dependence, we need to analyze these two data sets simultaneously.
To that end, we perform a combined fit of the \gls{nlo} \gls{iam} to the $\pi\pi$ lattice data of Ref.~\cite{Alexandrou:2017mpi} and Ref.~\cite{Dudek2013}, the former being associated with the $\SI{317}{\MeV}$ $\gamma\pi$ data and the latter with the $\SI{391}{\MeV}$ ones.
Since the two sets were independently generated, the $\chi^2$ is the sum of two terms in the form of \Eqref{eq:chi_2_pi_pi}, one for each data set.
A graphical comparison of the fit result with the data is shown in \Figref{fig:iam_fit}, while the goodness of the fit is shown in \Tabref{table:pi_pi_fit} together with the obtained value of $l^r$, the resulting $p$-value of $\SI{20}{\percent}$ being reasonable.
There is a $2\sigma$ tension with the \gls{chpt} value of $l^r$, however, this deviation comes at no surprise, given the unitarization via the \gls{iam}~\cite{Dobado:1992ha,Dobado:1996ps,Guerrero:1998ei,GomezNicola:2001as}.
With $l^r$ fixed, we continue the $\pi\pi$ $P$-wave via \Eqref{eq:pi_pi_second_sheet} to the second Riemann sheet to determine the $\rho$ characteristics at the physical point as listed in \Tabref{table:pi_pi_fit}.
Comparing with the literature values given ibidem, we note a $4\sigma$ discrepancy in $\Mrho$ and a $2\sigma$ tension in $\Im(g_{\rho\pi\pi})$, while both the width and the real part of the coupling agree well.
This is explained by the fact that the \gls{nlo} \gls{iam} has only a single free parameter, leading to a trade-off between the different $\rho$ properties.
To improve on this, the \gls{nnlo} \gls{iam} can be employed~\cite{Niehus:2020gmf}, however, with data at only two different pion masses both exceeding $\SI{300}{\MeV}$, we find that stable fits are not feasible.

\subsection[Fits to $\gamma\pi$ data]{Fits to $\boldsymbol{\gamma\pi}$ data}
Next, we fit the $\gamma\pi$ data.
Since the pion-mass dependence of each fit parameter $b_{kj}$ is described by two free parameters, compare \Eqref{eq:subtraction_mass_dependence}, we can perform the fits to the two $\gamma\pi$ data sets independently, the fit parameters being the values of $b_{kj}$ at the two different pion masses.
Hence at this stage we can work in lattice units, with different units for each data set.
\begin{table}
  \begin{center}
    \begin{tabular}{lllll}
      \toprule
      & & Ref.~\cite{Alexandrou:2018jbt} & Refs.~\cite{Briceno:2015dca, Briceno:2016kkp} & combined \\
      \midrule
      $\ROMAN{1}$ & $\frac{\chi^2}{\text{dof}}$ & $\frac{57.8 }{ 48 - 5} = 1.34$ & $\frac{67.0 }{ 37 - 5} = 2.09$ & $\frac{57.8 + 67.0}{85 - 10} = 1.66$ \\
                  & $p$-value & $\num{6.54e-2}$ & $\num{2.81e-4}$ & $\num{2.70e-4}$ \\
      \midrule
    $\ROMAN{1}\mscr{P}$ & $\frac{\chi^2}{\text{dof}}$ & $\frac{61.1 }{ 48 - 5} = 1.42$ & $\frac{44.0 }{ 37 - 5} = 1.37$ & $\frac{61.1 + 44.0}{85 - 10} = 1.40$ \\
      & $p$-value & $\num{3.61e-02}$ & $\num{7.70e-02}$ & $\num{1.26e-02}$ \\
      \midrule
      $\ROMAN{2}$ & $\frac{\chi^2}{\text{dof}}$ & $\frac{59.2 }{ 48 - 5} = 1.38$ & $\frac{53.9 }{ 37 - 5} = 1.69$ & $\frac{59.2 + 53.9}{85 - 10} = 1.51$ \\
      & $p$-value & $\num{5.13e-2}$ & $\num{8.99e-3}$ & $\num{2.96e-3}$ \\
      \midrule
      $\ROMAN{2}\mscr{P}$ & $\frac{\chi^2}{\text{dof}}$ & $\frac{57.9 }{ 48 - 5} = 1.35$ & $\frac{43.6 }{ 37 - 5} = 1.36$ & $\frac{57.9 + 43.6}{85 - 10} = 1.35$ \\
      & $p$-value & $\num{6.43e-02}$ & $\num{8.31e-2}$ & $\num{2.26e-2}$ \\
      \midrule
      $\ROMAN{3}$ & $\frac{\chi^2}{\text{dof}}$ & $\frac{59.5 }{ 48 - 5} = 1.38$ & $\frac{51.9 }{ 37 - 5} = 1.62$ & $\frac{59.5 + 51.9}{85 - 10} = 1.49$ \\
      & $p$-value & $\num{4.83e-2}$ & $\num{1.44e-2}$ & $\num{4.04e-3}$ \\
      \midrule
      $\ROMAN{3}\mscr{P}$ & $\frac{\chi^2}{\text{dof}}$ & $\frac{57.2 }{ 48 - 5} = 1.33$ & $\frac{43.6 }{ 37 - 5} = 1.36$ & $\frac{57.2 + 43.6}{85 - 10} = 1.34$ \\
      & $p$-value & $\num{7.20e-2}$ & $\num{8.32e-2}$ & $\num{2.51e-2}$ \\
      \bottomrule
    \end{tabular}
    \caption{The quality of the fit to the $\gamma\pi$ data for the different parameterizations of the subtraction functions.}\label{table:gamma_pi_fit}
  \end{center}
\end{table}
We use $n=2$ subtractions in \Eqref{eq:partial_wave}, for once subtracted \gls{kt} equations fail to describe the energy dependence of the data correctly and thus do not allow for statistically acceptable fits.
Note that increasing the number of subtractions to $n=3$ does not provide additional flexibility, since the reconstruction theorem~\eqref{eq:reconstruction_theorem} is invariant under the shift $\mc{B}(s, q^2) \mapsto \mc{B}(s, q^2) + \lambda(q^2)(3s - 3\Mp^2 - q^2)$ with $\lambda$ an arbitrary function, hence one subtraction function can be eliminated.
This shift is forbidden for $n=2$ due to the high-energy behavior of $\mc{B}$, but becomes possible for $n=3$.
In addition, we pick $N=2$, that is, we have three fit parameters $b_{kj}$ in $c_0$ and two in $c_1$.
If instead $N=1$ is used, the fit quality becomes poor, while at $N=3$ the fit stability deteriorates.
The exception are the strategies $\ROMAN{3}$ and $\ROMAN{3}\mscr{P}$, where we pick $N=3$, which again amounts to five fit parameters due to \Eqref{eq:modified_threshold}.

\begin{figure}
  \begin{center}
   \input{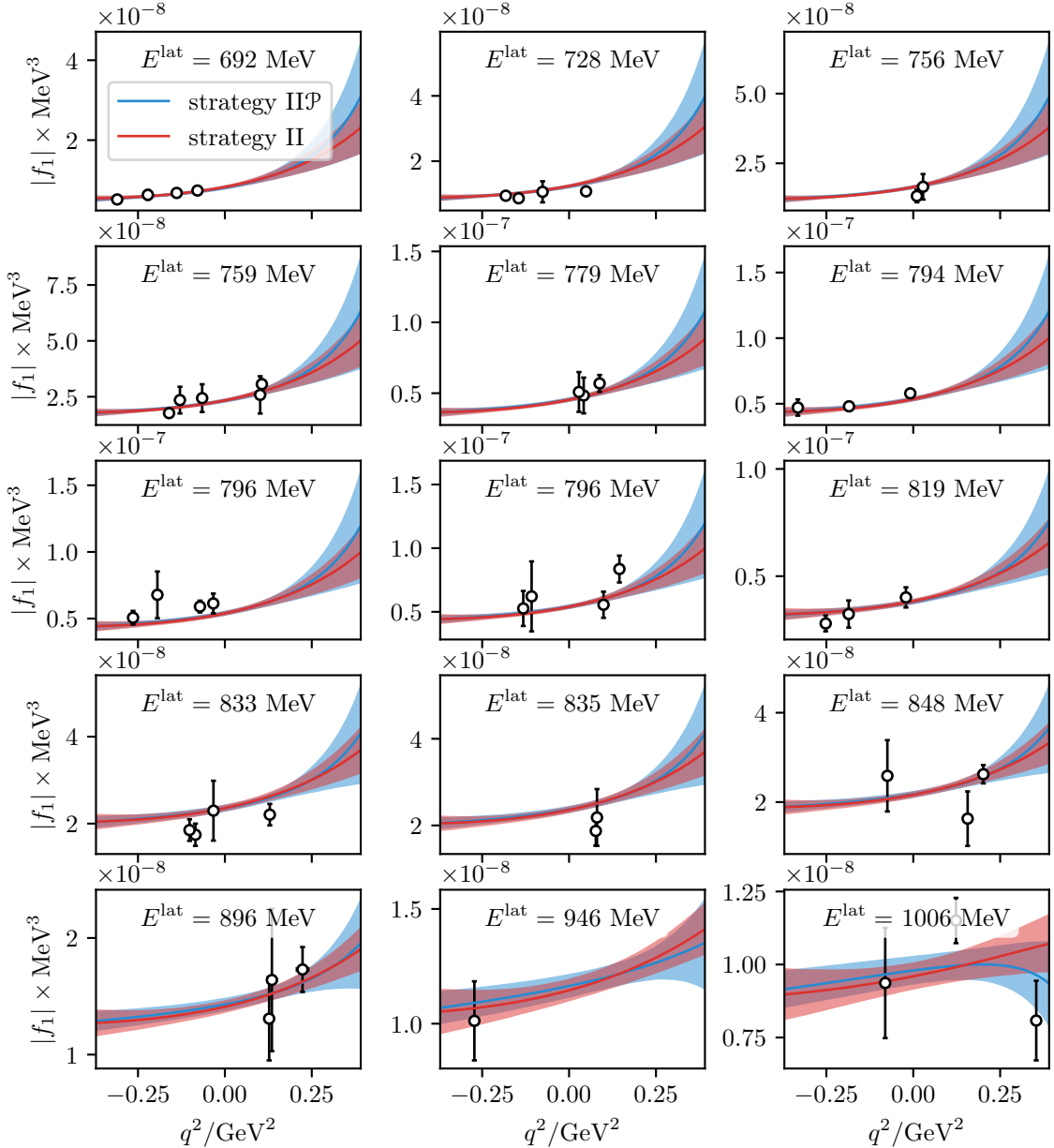}
    \caption{The results of two fit strategies in comparison with the $\gamma\pi$ lattice data of Ref.~\cite{Alexandrou:2018jbt} at $\Mp\approx\SI{317}{\MeV}$. Shown are slices of constant energy. For convenience, the results are displayed in physical units, but the fit is carried out in lattice units, thus the error bands represent the statistical error only.}\label{fig:gamma_pi_data_320}
  \end{center}
\end{figure}

\begin{figure}
  \begin{center}
    \input{img/gamma_pi_data_390.pgf}
    \caption{As \Figref{fig:gamma_pi_data_320}, but for the lattice data of Refs.~\cite{Briceno:2015dca,Briceno:2016kkp} at $\Mp\approx\SI{391}{\MeV}$.}\label{fig:gamma_pi_data_390}
  \end{center}
\end{figure}

To obtain statistically acceptable fits to the data at $\Mp \approx \SI{391}{\MeV}$, we need to exclude the six data points at the highest energy, $E^\text{lat}\approx\SI{1096}{\MeV}$.
These points lie far above the resonance region, for although at this pion mass the $\rho$ is heavy, i.e.,  $\Mrho = 846.1(3.1)(3.2)(0.1)\,\si{\MeV}$ (errors as in \Tabref{table:pi_pi_fit}), its width $\Gammarho = 10.8(8)(9)(1)\,\si{\MeV}$ is tiny.
Moreover, several of the data points with the smallest absolute errors of $\abs{\mc{A}_\text{FV}}$ are located at this energy.
Hence the six excluded data points provide rather strong constraints on the asymptotic high-energy behavior of the \gls{kt} equations instead of the resonance physics in which we are primarily interested.

\begin{figure}
  \begin{center}
     \scalebox{0.995}{\input{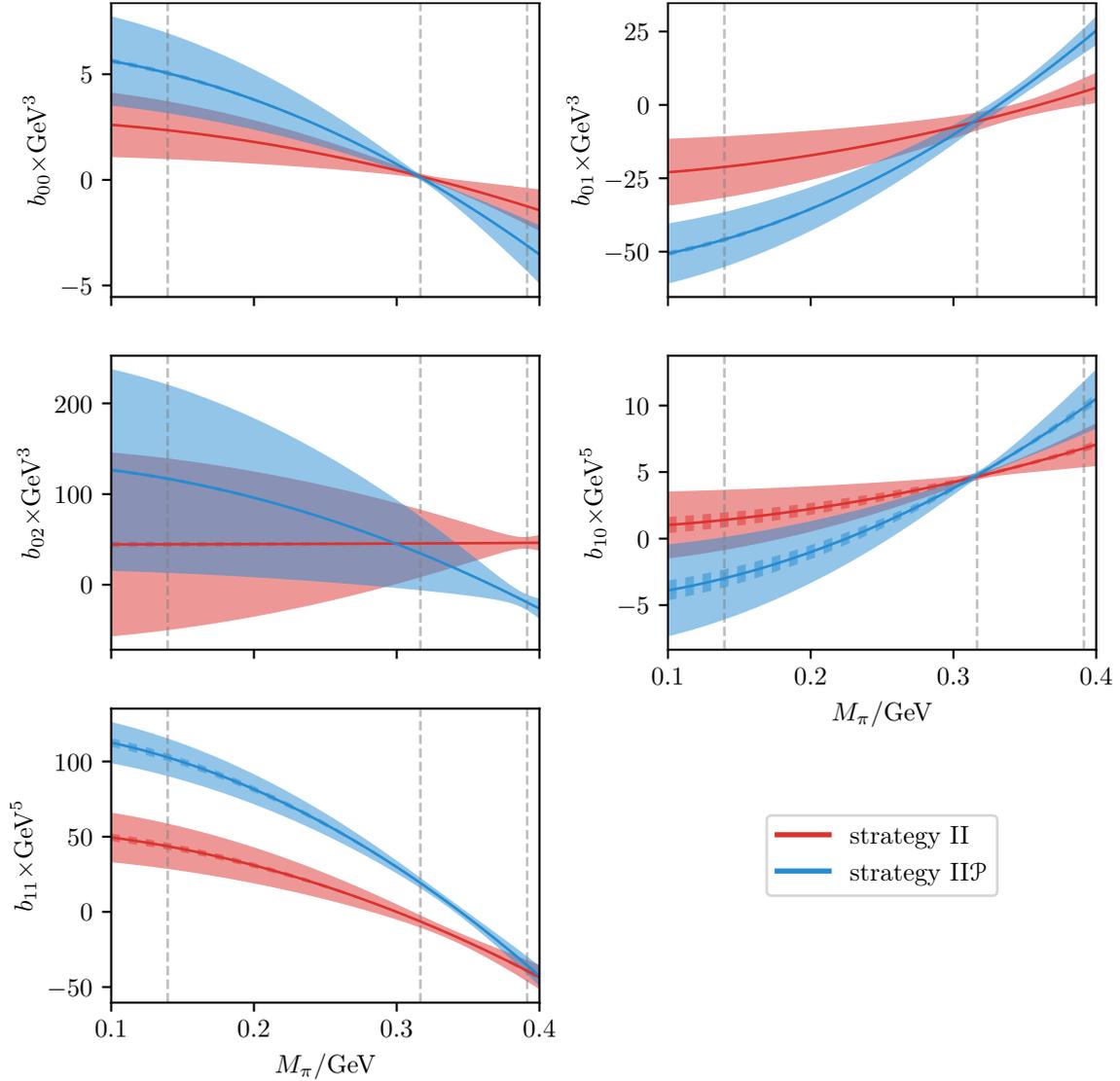}}
    \caption{The pion-mass dependence of the fit parameters for two different fit strategies. The dashed gray lines mark the physical pion mass and the ones of the two lattice data sets, the dashed error bands correspond to the error of the lattice spacings, while the filled ones are associated with the statistical error.}
    \label{fig:subtraction_constants}
  \end{center}
\end{figure}

We carry out fits for each strategy enumerated in \Tabref{tab:strategies}, with an overview of the fit qualities given in \Tabref{table:gamma_pi_fit}.
While the goodness of the fit at the lower pion mass is rather insensitive to the parameterization of the subtraction functions, the data at the higher pion mass is more selective, because the relative error of $\abs{\mc{A}_\text{FV}}$ at the higher pion mass is smaller than the error at lower mass.
Notably, we observe improvement when including a pole factor, this is true for all strategies, with the overall $p$-value improving by at least an order of magnitude in each case, and even by two orders of magnitude when going from strategy $\ROMAN{1}$ to $\ROMAN{1}\mscr{P}$.
As soon as a pole factor is included, it does not matter much if the remaining $q^2$-dependence is parameterized by a plain polynomial, a conformal one, or a conformal one with modified threshold behavior, the overall $p$-values of strategy $\ROMAN{1}\mscr{P}$, $\ROMAN{2}\mscr{P}$, and $\ROMAN{3}\mscr{P}$ are similar, with a slight improvement when using conformal parameterizations.
Hence in the following we group the results of the three parameterizations including a pole together.
If no pole is used, at higher pion mass the fit clearly disfavors a plain polynomial and instead prefers a conformal one, with only a very slight further improvement when modifying the threshold scaling.
Thus we exclude strategy $\ROMAN{1}$ and combine strategy $\ROMAN{2}$ and $\ROMAN{3}$.
As a representative of each group, we pick strategy $\ROMAN{2}$ and $\ROMAN{2}\mscr{P}$.
The corresponding partial waves are compared with the two lattice data sets in \Figref{fig:gamma_pi_data_320} and \Figref{fig:gamma_pi_data_390}.
As can be observed, independently of the presence of a pole factor, the magnitude of $f_1$ increases with growing $q^2$, in accordance with phenomenology~\cite{Hoferichter:2014vra}.
To check if we are sensitive to the mixed rescattering effects included in the \gls{kt} equations, we re-perform the fits with the replacement $\mc{B}_k(s, q^2) \mapsto s^k\Omega(s)$.
At $\Mp\approx\SI{317}{\MeV}$ we obtain a $p$-value of $\num{4.97e-2}$ with strategy $\ROMAN{2}$ and $\num{6.26e-2}$ with strategy $\ROMAN{2}\mscr{P}$, while at $\Mp\approx\SI{391}{\MeV}$ we obtain $\num{8.77e-3}$ and $\num{8.49e-2}$, respectively.
Comparing with the corresponding entries of \Tabref{table:gamma_pi_fit}, the observed difference is insignificant, thus we conclude that mixed rescattering does not need to be taken into account to describe the data at the present level of precision.

\begin{figure}
  \begin{center}
    \input{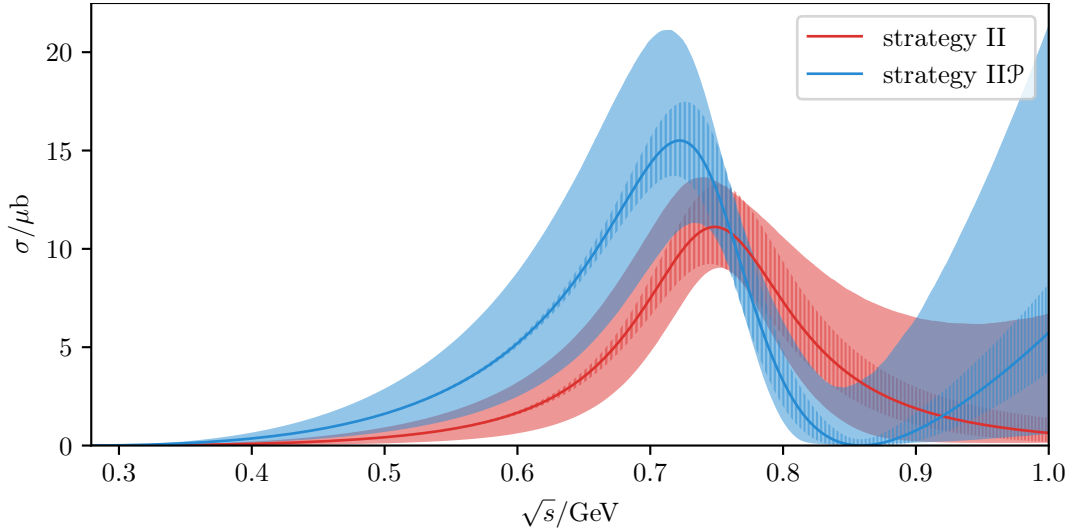}
    \caption{The cross section at the physical point for two different fit strategies. The error bands are as in \Figref{fig:subtraction_constants}.}\label{fig:cross_section}
  \end{center}
\end{figure}

\subsection{Chiral extrapolation}
Equipped with the \gls{kt} fit results, we can determine the pion-mass dependence of the fit parameters via \Eqref{eq:subtraction_mass_dependence}.
To that end, we need to translate the fit parameters associated with the two different data sets to a common set of units, hence the errors of the lattice spacings enter the picture.
Since the scale of the two data sets is set in different ways (via the $\Upsilon(1S)$--$\Upsilon(2S)$ splitting at $\Mp\approx\SI{317}{\MeV}$ and via the $\Omega$ baryon mass at $\Mp\approx\SI{391}{\MeV}$), an additional systematic error arises, which is difficult to quantify (this also applies to the fit of the $\pi\pi$ data).
However, compared to the sizable statistical uncertainty of the data and the systematic error of the chiral extrapolation to be discussed in Sec.~\ref{sec:chiral_anomaly_etc}, the systematic error associated with the scale setting is likely irrelevant at present. Therefore, the uncertainty associated with the lattice spacing given in the remainder of this work will always refer  to its statistical error only.\footnote{There is also a systematic error from the continuum extrapolation, since the calculation in Ref.~\cite{Alexandrou:2018jbt} was performed at a single lattice spacing of $a\approx 0.11\fm$, while the one in Ref.~\cite{Briceno:2016kkp} was performed on an anisotropic lattice with temporal spacing $a_t\approx 0.03\fm$ and spatial spacing $a_s\approx 0.12\fm$.}
The pion-mass dependence of the fit parameters is depicted in \Figref{fig:subtraction_constants}.
While the leading parameters in the series expansion~\eqref{eq:conformal_polynomial}, $b_{00}$ and $b_{10}$, are constrained more strongly by the data at lower pion mass than the one at higher pion mass, the opposite is true for the highest-order term associated with $b_{02}$.
The latter comes at no surprise, for the data at higher mass contain much larger virtualities in the spacelike region, exceeding in absolute value the timelike virtualities of both data sets significantly, giving thus more weight to the $b_{02}$ term.
With decreasing pion mass, the $\omega$ pole moves from the real axis below the $3\pi$ threshold on the first Riemann sheet into the complex plane on the second sheet.
Since the pole factor $\mscr{P}$ that is present in strategies $\ROMAN{1}\mscr{P}$, $\ROMAN{2}\mscr{P}$, and $\ROMAN{3}\mscr{P}$ describes a bound state, naturally the question arises if the change in the nature of the pole needs to be reflected in the extrapolation in the pion mass for these strategies. A resonant $\omega$ could be implemented via a dispersively-improved Breit--Wigner parameterization, which, in practice, is almost indistinguishable from a pole ansatz unless very close to the singularities.
Given the large uncertainties of $b_{jk}$ at the physical point, this change is thus immaterial, especially, since for the extraction of the observables the subtraction functions need to be evaluated at vanishing virtuality only, and for every strategy $c_k(0) = b_{k0}$ holds.
Taking care of the pion-mass dependence of the fit parameters, the \gls{iam}, and the \gls{kt} equations, we can extrapolate the partial wave to the physical point.
Computing the cross section via \Eqref{eq:cross_section} yields the line shape shown in \Figref{fig:cross_section} exhibiting the characteristic resonance peak.
In both fit strategies, the error increases when moving beyond the resonance, which reflects the fact that most data points lie around the resonance region.
In principle, the omitted data points at the highest energy could provide further constraints, but since no acceptable fits could be found when including these points, we conclude that with the currently available lattice data the asymptotic form of the cross section remains largely unconstrained. 
In this regard, we remark that the \gls{kt} basis functions with $n=2$ subtractions increase too fast asymptotically compared to expectations from the Froissart bound~\cite{Froissart:1961ux}, so that a proper high-energy completion needs to be imposed~\cite{Hoferichter:2017ftn}. However, these considerations become relevant only well beyond $1\GeV$ and thus do not affect the current fit, for which the $n=2$ subtraction scheme provides the adequate number of free parameters to be able to describe both the chiral anomaly and the $\rho$-meson properties~\cite{Hoferichter:2017ftn}.

\subsection{Chiral anomaly and radiative coupling}\label{sec:chiral_anomaly_etc}
Finally, we can determine the anomaly $F_{3\pi}$ and the radiative coupling at the physical point via \Eqref{matching} and \Eqref{eq:partial_wave_residue}, respectively.
The values are listed for the different fit strategies in \Tabref{table:anomaly_and_coupling}.
Since the outcomes of the different fit variants are highly correlated with only minor differences in fit quality and very similar statistical errors, we do not compute weighted averages, but instead only perform plain averages to determine the central values.
Doing so for the acceptable fits without a pole factor, i.e., averaging over strategy $\ROMAN{2}$ and $\ROMAN{3}$, results in
\begin{equation}\label{eq:observables_without_pole}
  \begin{split}
    F_{3\pi} &= 24(13)(1)\,\si{\GeV^{-3}}, \\
    g_{\rho\gamma\pi} &= \left[0.51(6)(4) + i 0.03(13)(2)\right]\,\si{\GeV^{-1}},\\
    |g_{\rho\gamma\pi}| &= 0.51^{+0.08}_{-0.05}(4)\,\si{\GeV^{-1}},
  \end{split}
\end{equation}
with errors as in \Tabref{table:anomaly_and_coupling}, while the strategies including an $\omega$ pole, i.e., $\ROMAN{1}\mscr{P}$, $\ROMAN{2}\mscr{P}$, and $\ROMAN{3}\mscr{P}$, yield
\begin{equation}\label{eq:observables_with_pole}
  \begin{split}
  F_{3\pi} &= 47(18)(1)\,\si{\GeV^{-3}}, \\
  g_{\rho\gamma\pi} &= \left[0.60(8)(4) + i 0.26(18)(3)\right]\,\si{\GeV^{-1}},\\
  |g_{\rho\gamma\pi}| &= 0.66^{+0.15}_{-0.12}(3)\,\si{\GeV^{-1}}.
  \end{split}
\end{equation}
Both values of $F_{3\pi}$ are compatible with the prediction~\eqref{eq:anomaly}, albeit only due to their large errors.
Fits including the pole ansatz do display a better fit quality, but not at a level that would conclusively demonstrate the necessity of the pole. Since, further, both fit variants agree within statistical uncertainties, we conclude that the current lattice data cannot discriminate between \Eqref{eq:observables_without_pole} and \Eqref{eq:observables_with_pole} and quote the resulting spread as an additional systematic error.  
This error also arises due to the absence of lattice data at several different pion masses by one collaboration, forcing us to fit our representation to two data sets by two different collaborations at only two different pion masses, which makes it impossible to fix the pion-mass dependence of the subtraction functions beyond the simple ansatz~\eqref{eq:subtraction_mass_dependence}.
Averaging over all fit results except for strategy $\ROMAN{1}$, we finally quote
\begin{equation}
  \begin{split}
  F_{3\pi} &= 38(16)(1)(11)\,\si{\GeV^{-3}}, \\
  g_{\rho\gamma\pi} &= \left[0.57(7)(4)(4) + i 0.17(16)(3)(12)\right]\,\si{\GeV^{-1}},\\
  |g_{\rho\gamma\pi}| &= 0.60^{+0.12}_{-0.09}(3)(7)\,\si{\GeV^{-1}},
  \end{split}
  \label{eq:final}
\end{equation}
where the last error is our estimate of the systematic uncertainty associated with the parameterization of the subtraction functions. 

\begin{table}
  \begin{center}
    \begin{tabular}{llll}
      \toprule
      & $F_{3\pi}\times\si{\GeV^3}$ & $\Re(g_{\rho\gamma\pi})\times\si{\GeV}$ & $\Im(g_{\rho\gamma\pi})\times\si{\GeV}$ \\
      \midrule
      $\ROMAN{1}$ & $13(11)(0)$ & $0.50(6)(4)$ & $0.09(11)(2)$ \\
      $\ROMAN{1}\mscr{P}$ & $46(18)(1)$ & $0.59(8)(4)$ & $0.26(18)(3)$ \\
      $\ROMAN{2}$ & $23(13)(1)$ & $0.51(6)(4)$ & $0.02(13)(2)$ \\
      $\ROMAN{2}\mscr{P}$ & $48(18)(1)$ & $0.61(8)(4)$ & $0.27(18)(3)$ \\
      $\ROMAN{3}$ & $26(13)(1)$ & $0.52(6)(4)$ & $0.05(13)(2)$ \\
      $\ROMAN{3}\mscr{P}$ & $48(18)(1)$ & $0.61(8)(4)$ & $0.27(18)(3)$ \\
      \bottomrule
    \end{tabular}
    \caption{The anomaly and the radiative coupling at the physical point. The fit uncertainty gives the first error, the second error corresponds to the error of the lattice spacings.}\label{table:anomaly_and_coupling}
  \end{center}
\end{table}

The resulting value of $F_{3\pi}$ is perfectly consistent with the chiral prediction~\eqref{eq:anomaly}, but carries a large uncertainty. This is the first extraction of this low-energy parameter from lattice-QCD calculations, and will improve accordingly once better data become available. The residue $g_{\rho\gamma\pi}$ is currently not known better than from an $SU(3)$ \gls{vmd} estimate~\cite{Klingl:1996by}, which suggests $|g_{\rho\gamma\pi}|=0.79(8)\si{\GeV^{-1}}$~\cite{Hoferichter:2017ftn}, again compatible with \Eqref{eq:final} (within $1.2\sigma$).\footnote{The branching fractions cited in Ref.~\cite{ParticleDataGroup:2020ssz} imply $|g_{\rho\gamma\pi}|=0.72(4)\,\si{\GeV^{-1}}$ for the charged channel and $|g_{\rho\gamma\pi}|=0.73(6)\,\si{\GeV^{-1}}$ for the neutral one. However, these values derive from high-energy Primakoff measurements~\cite{Jensen:1982nf,Huston:1986wi,Capraro:1987rp} and \gls{vmd} fits to $e^+e^-\to \pi^0\gamma$ data~\cite{CMD-2:2004ahv,SND:2016drm}, respectively, and thus involve a substantial model dependence.} The difference to the \gls{vmd} estimate increases to $2.3\sigma$ for Eq.~\eqref{eq:observables_without_pole}, while there is full agreement with Eq.~\eqref{eq:observables_with_pole}.
This provides a-posteriori evidence for the presence of an $\omega$ pole in the subtraction functions, as does the final result for the cross section shown in Fig.~\ref{fig:cross_section} when compared to the expected peak cross section around $20\,\mu\text{b}$~\cite{Seyfried:2017}. 
The radiative coupling has also been extracted in Ref.~\cite{Alexandrou:2018jbt} under the assumption that the pion-mass dependence of $|G_{\rho\gamma\pi}|=|g_{\rho\gamma\pi}|\Mp/2$ is weak, leading to $|g_{\rho\gamma\pi}|_{\text{\cite{Alexandrou:2018jbt}}}=1.15(5)(3)\,\si{\GeV^{-1}}$. 
This value differs from the \gls{vmd} estimate by $3.6\sigma$, a discrepancy that
went unnoticed in Ref.~\cite{Alexandrou:2018jbt} because it is mitigated 
by a missing factor $2$ in Eq.~(17) for $\Gamma(\rho\to\pi\gamma)$ therein~\cite{Petschlies:2021}. 
Moreover,  our analysis shows that the uncertainties especially from the chiral extrapolations are substantially larger. In particular, a pion-mass independent $|G_{\rho\gamma\pi}|$ renders the residue divergent in the chiral limit, while at $\Mp=317\,\si{MeV}$ one has $|g_{\rho\gamma\pi}|_{\text{\cite{Alexandrou:2018jbt}}}^{\Mp=317\,\si{MeV}}=0.507(20)(13)\,\si{\GeV^{-1}}$ as well as $\vert g_{\rho\gamma\pi}\vert^{\Mp=317\,\si{MeV}}=0.552(18)(18)(0)$, the latter being the average~\eqref{eq:final} at this pion mass. 
We conclude that $|g_{\rho\gamma\pi}|$ instead of $|G_{\rho\gamma\pi}|$ is approximately pion-mass independent, thus avoiding the divergence in the chiral limit.

\section{Conclusions}\label{sec:conclusions}
In this work we analyzed state-of-the-art $\gamma\pi\to\pi\pi$ lattice-\gls{qcd} data using a combination of dispersion relations and \gls{chpt}, to be able to describe both the momentum and the pion-mass dependence in a reliable manner. 
Extrapolating to the physical point, we determined the cross section, extracted the radiative coupling of the $\rho$ meson, and, for the first time, the chiral anomaly $F_{3\pi}$, see \Eqref{eq:final} for the final results, and \Eqref{eq:observables_with_pole} for a variant that imposes the $\omega$ pole in the subtraction functions.
These results agree with expectations from \gls{chpt} and phenomenology, albeit within large uncertainties. 
By combining \gls{kt} equations with the pion-mass dependence as described by the \gls{iam}, we could thus confront predictions from the \gls{kt} framework with lattice \gls{qcd}, emphasizing the role of the reaction $\gamma\pi\to\pi\pi$ to develop methods that could subsequently be applied to more complicated processes. 
While the current lattice data are not yet sensitive to the mixed rescattering effects included in \gls{kt} equations, this work demonstrates that the framework 
is not only a valuable tool for the description of experimental data, but that it also applies to the analysis of lattice calculations.
Future lattice-\gls{qcd} computations are expected to reduce the statistical uncertainties, hence the data will be able to differentiate between parameterizations of the pion-mass dependence of the subtraction functions, which currently represents the largest source of systematic uncertainties, e.g., with the presence of an $\omega$ pole in the subtraction functions preferred by comparison to phenomenology, but not resolved within the lattice calculations. 
Moreover, once data at more different pion masses become available, more refined parameterizations can be employed, and if such data are analyzed with a single scale-setting strategy, this will remove the systematic error stemming from the simultaneous use of different ways to set the scale.
Such refined lattice-\gls{qcd} calculations, in combination with the analysis tools developed in this work, have the potential to improve several important low-energy parameters of \gls{qcd}, most notably the chiral anomaly $F_{3\pi}$, complementary to future experimental determinations, e.g., from the COMPASS Primakoff program~\cite{Seyfried:2017}.   
We also expect that the strategy pursued here, combining dispersion relations with effective-field-theory methods for the pion-mass dependence, will have broad applications in particular to other processes dominated by $\pi\pi$ dynamics.

\begin{acknowledgments}
The lattice data taken from Refs.~\cite{Dudek2013,Briceno:2015dca,Briceno:2016kkp} were provided by the \glsfirst{hadspec}---no endorsement on their part of the analysis presented in the current paper should be assumed.
In addition, we would like to thank the authors of Refs.~\cite{Alexandrou:2017mpi,Alexandrou:2018jbt} for sharing lattice data with us.
We thank Marcus Petschlies and Ra\'ul Brice\~no for valuable discussions, and Christopher Thomas for comments on the manuscript.
Financial support by the Bonn--Cologne Graduate School of Physics and Astronomy (BCGS), the DFG (CRC 110, ``Symmetries and the Emergence of Structure in QCD''), and the Swiss National Science Foundation, under Project No.\ PCEFP2\_181117, is gratefully acknowledged.                          
\end{acknowledgments}

\appendix

\section{Kernel method}\label{sec:kernel_method}
Here we describe a way to facilitate the computation of the hat function $\hat{\mc{B}}_k$ as defined in \Eqref{eq:khuri_treiman} and needed in the evaluation of the partial wave via \Eqref{eq:partial_wave}.
To that end, we extend the discussion given in Refs.~\cite{Hoferichter:2017ftn,Dax:2020dzg} to non-vanishing virtualities.
The starting point is the simple dispersive representation
\begin{equation}\label{eq:spectral_representation}
  \mc{B}_k\left(s, q^2\right)
  = P_k\left(s\right) + \frac{s^d}{\pi}\int\limits_{4\Mp^2}^\infty\frac{\Im\left[\mc{B}_k\left(x, q^2\right)\right]}{\left(x-s\right)x^d}\diff x,
\end{equation}
with $d$ the number of subtractions and $P_k = \sum_{j=0}^{d-1} h_{kj} s^j$ the subtraction polynomial whose real coefficients $h_{kj}$ can be determined via matching to \Eqref{eq:khuri_treiman}.
For instance, in the case of interest $\delta = \pi$ for large values of $s$, hence $\Omega(s) = O(1/s)$, accordingly, \Eqref{eq:khuri_treiman} implies $\mc{B}_k(s, q^2) = O(s^{n-2})$.
Given $n=2$ it is thus more than sufficient to use $d=2$.
Equating the Taylor expansions around $s=0$ of both \Eqref{eq:spectral_representation} and the definition of $\mc{B}_k$ in \Eqref{eq:khuri_treiman} we obtain
\begin{equation}\label{eq:kernel_polynomial_coefficients}
  h_{00} = 1, \quad
  h_{01} = \frac{\diff \Omega}{\diff s}\left(0\right), \quad
  h_{10} = 0, \quad
  h_{11} = 1,
\end{equation}
where we used $\Omega(0) = 1$.
Coming back to the general case, inserting \Eqref{eq:spectral_representation} into the definition of $\hat{\mc{B}}_k$ in \Eqref{eq:khuri_treiman} yields
\begin{equation}\label{eq:dispersive_integral_via_kernel}
  \hat{\mc{B}}_k\left(s, q^2\right)
  = \sum_{j=0}^{d-1} h_{kj}G_j\left(s, q^2\right) + \frac{1}{\pi}\int\limits_{4\Mp^2}^\infty W_d\left(s, q^2, x\right)\Im\left[\mc{B}_k\left(x, q^2\right)\right]\diff x,
\end{equation}
with
\begin{equation}
  \begin{split}
  G_j\left(s, q^2\right) &= 6\sum_{i=0}^{\lfloor j/2 \rfloor}{j\choose 2i}\frac{\tau\left(s, q^2\right)^{j-2i}\kappa\left(s, q^2\right)^{2i}}{\left(2i+1\right)\left(2i+3\right)}, \\
  W_d\left(s, q^2, x\right) &=\frac{3}{2x^d}\sum_{k=0}^d{d\choose k}\tau\left(s, q^2\right)^{d-k}\kappa\left(s, q^2\right)^{k-1}\int\limits_{-1}^1\frac{z^k\left(1-z^2\right)}{\xi\left(s, q^2, x\right) - z}\diff z,
  \end{split}
\end{equation}
where we decomposed Mandelstam $t$ according to \Eqref{eq:mandelstam_t} and introduced
\begin{equation}
  \xi\left(s, q^2, x\right)
  = \frac{x - \tau\left(s, q^2\right)}{\kappa\left(s, q^2\right)}.
\end{equation}
For low $d$ the kernel $W_d$ simplifies to~\cite{Hoferichter:2017ftn,Dax:2020dzg}
\begin{equation}\label{eq:kernel_simple_cases}
  W_1\left(s, q^2, x\right) = \mc{W}\left(s, q^2, x\right) - \frac{2}{x}, \quad
  W_2\left(s, q^2, x\right) = W_1\left(s, q^2, x\right) - \frac{2\tau\left(s, q^2\right)}{x^2},
\end{equation}
with
\begin{equation}\label{eq:kernel_core}
  \begin{split}
  \mc{W}\left(s, q^2, x\right)
  &= \frac{3}{2\kappa\left(s,q^2\right)}\int\limits_{-1}^1\frac{1-y^2}{\xi\left(s,q^2,x\right)-y}\diff y \\
  &= \frac{3}{\kappa\left(s, q^2\right)}\left[\left(1-\xi\left(s,q^2, x\right)^2\right)Q_0\left(\xi\left(s,q^2, x\right)\right) + \xi\left(s,q^2, x\right)\right],
  \end{split}
\end{equation}
where
\begin{equation}
  Q_0\left(\xi\right)
  = \frac{1}{2}\int\limits_{-1}^1\frac{1}{\xi-y}\diff y
  = \frac{1}{2}\left[\log\left(\frac{1+\xi}{1-\xi}\right) -i\pi\Sign\left(\Im\left(\xi\right)\right)\right]
\end{equation}
is the lowest Legendre function of the second kind.
Here we use the principal branch of the logarithm with a cut along the negative real axis.
If $s = 4\Mp^2$ or, in case of timelike virtualities, $s = (\sqrt{q^2} + \Mp)^2$, $\kappa(s, q^2)$ vanishes.
Equation~\eqref{eq:kernel_core} implies that $\mc{W}$ has singularities whenever this happens.
To prove that these are removable and to derive a representation that is suitable for numerical implementation, we note that $\kappa\to 0$ implies $\xi\to\infty$, hence we expand $Q_0$ around $\xi^{-1} = 0$, arriving at
\begin{equation}\label{eq:partial_wave_kernel_core_series}
  \mc{W}\left(s, q^2, x\right)
  = \frac{6}{x-\tau\left(s\right)}\sum_{j=0}^\infty\frac{1}{\left(2j+1\right)\left(2j+3\right)}\left(\frac{1}{\xi\left(s,q^2,x\right)}\right)^{2j}.
\end{equation}
According to the ratio test, \Eqref{eq:partial_wave_kernel_core_series} converges absolutely as long as $\vert\xi\vert > 1$.
In particular, we see that $\lim_{\xi\to\infty}\mc{W}$ is manifestly finite.
In passing, we note that the terms subtracted from $\mc{W}$ in \Eqref{eq:kernel_simple_cases} precisely cancel out the leading terms in \Eqref{eq:partial_wave_kernel_core_series} such that $W_d(s, q^2, x) = O(x^{d + 1})$, as it needs to be the case to ensure convergence of the integral in \Eqref{eq:dispersive_integral_via_kernel}.
Applying \Eqref{eq:dispersive_integral_via_kernel} in the case of interest we obtain
\begin{equation}\label{eq:kernel_method_hat_function_final}
  \hat{\mc{B}}_k\left(s, q^2\right)
  = 2 h_{k0} + 2\tau\left(s, q^2\right) h_{k1} + \frac{1}{\pi}\int\limits_{4\Mp^2}^\infty W_2\left(s, q^2, x\right)\Im\left[\mc{B}_k\left(x, q^2\right)\right]\diff x,
\end{equation}
with the coefficients $h_{kj}$ given in \Eqref{eq:kernel_polynomial_coefficients} and $W_2$ to be computed via Eqs.~\eqref{eq:kernel_simple_cases},~\eqref{eq:kernel_core}, and~\eqref{eq:partial_wave_kernel_core_series}.
Clearly, \Eqref{eq:kernel_method_hat_function_final} allows for evaluation of the hat function at arbitrary complex values of $s$ requiring the imaginary part of $\mc{B}_k$ along the physical scattering region only.
Because the high-energy region in the integral in \Eqref{eq:kernel_method_hat_function_final} is strongly suppressed by the asymptotic behavior of $W_2$, the integral can be cut off at high energies.

\bibliography{literature} 
\end{document}